\pdfoutput=1
\PassOptionsToPackage{obeyspaces}{url}
\documentclass[twocolumn,prb,a4paper,final,superscriptaddress,longbibliography]{revtex4-1}
\usepackage{amsmath,amssymb}
\usepackage{graphicx}
\usepackage{comment}
\usepackage{bm}
\usepackage[colorlinks=true,allcolors=blue]{hyperref}
\usepackage{color}
\usepackage{subfigure}
\newcommand{\Eprint}[2]{\href{#1}{\urlstyle{same}\nolinkurl{#2}}}

\newcommand{\pd}{\partial}

\newcommand{\sign}{\qopname\relax o{sign}}

\newcommand{\const}{\text{const}}

\newcommand{\integers}{\mathbb{Z}}

\newcommand{\arccot}{\qopname\relax o{arccot}}
\newcommand{\am}{\qopname\relax o{am}}
\newcommand{\K}{\qopname\relax o{K}}
\newcommand{\E}{\qopname\relax o{E}}

\newcommand{\calH}{{\cal H}}
\newcommand{\n}{{\bf n}}
\newcommand{\D}{{\bf D}}
\newcommand{\Tkappa}{\tilde{\kappa}}

\begin{document}
\title{Domain-wall skyrmion chain
and domain-wall bimerons
in chiral magnets}
	
\date{\today}
\author{Yuki Amari}
\affiliation{Research and Education Center for Natural Sciences, Keio University, Hiyoshi 4-1-1, Yokohama, Kanagawa 223-8521, Japan}
\affiliation{Department of Physics, Keio University, 4-1-1 Hiyoshi, Kanagawa 223-8521, Japan}

\author{Calum Ross}
\affiliation{Research and Education Center for Natural Sciences, Keio University, Hiyoshi 4-1-1, Yokohama, Kanagawa 223-8521, Japan}
\affiliation{Department of Computer Science, Edge Hill University, St Helens Rd., Ormskirk L39 4QP, UK}

\author{Muneto Nitta}
\affiliation{Research and Education Center for Natural Sciences, Keio University, Hiyoshi 4-1-1, Yokohama, Kanagawa 223-8521, Japan}
\affiliation{Department of Physics, Keio University, 4-1-1 Hiyoshi, Kanagawa 223-8521, Japan}
\affiliation{International Institute for Sustainability with Knotted Chiral Meta Matter(SKCM$^2$), Hiroshima University, 1-3-2 Kagamiyama, Higashi-Hiroshima, Hiroshima 739-8511, Japan}

\vspace{.5in}
	
\begin{abstract}
We construct domain-wall skyrmion chains and domain-wall bimerons in chiral magnets with an out-of-plane easy-axis anisotropy and 
 without a Zeeman term coupling to a magnetic field.  Domain-wall skyrmions are skyrmions trapped inside a domain wall, they are present in the 
 ferromagnetic (FM) phase of a chiral magnet
with an out-of-plane easy-axis anisotropy.
In this paper, 
 we explore the stability of domain-wall skyrmions in the FM phase and in a chiral soliton lattice (CSL) or spiral phase, which is a periodic array of domain walls and anti-domain walls arranged in an alternating manner. 
In the FM phase,  
the worldline of a domain-wall skyrmion is bent to form a cusp at the position of the skyrmion. 
We describe such a cusp using both an analytic method 
and numerical solutions, 
and find a good agreement between them 
for small DM interactions.
We show that 
the cusp grows toward the 
phase boundary with the CSL, 
and eventually diverges at the boundary. 
Second, if we put one skyrmion trapped inside a domain wall in a CSL, it decays into a pair of merons by a reconnection of the domain wall 
and its adjacent anti-domain wall. 
Third, if we put skyrmions and anti-skyrmions alternately in domain walls and anti-domain walls,  respectively such a chain is stable. 
\end{abstract} 
\maketitle

%%%%%%%%%%%%%%%%%%%%%%%%%%%%%%%%%%%%%%%%%%%%%%%%%%%%%%%%%%%%%	
\section{Introduction}
\label{sec:intro}
Skyrmions are topologically stable field configurations first introduced by Skyrme as a model of nuclei \cite{Skyrme:1962vh}. They appear and have been studied in an array of physical models, including as the 
baryons in the large $N_{c}$ limit of 
quantum chromodynamics (QCD) \cite{Witten:1983tx}. More recently, most research on skyrmions has focused on magnetic skyrmions\cite{Bogdanov:1989,Bogdanov:1995} in chiral magnetic materials. These are a two-dimensional analog of nuclear skyrmions that are realized in chiral magnets with a Dzyaloshinskii-Moriya (DM) interaction \cite{Dzyaloshinskii,Moriya:1960zz}. Magnetic skyrmions have been observed in laboratory experiments \cite{doi:10.1126/science.1166767,doi:10.1038/nphys2045,doi:10.1038/nature09124} and are thought to have applications as information carriers in magnetic storage devices\cite{Nagaosa2013}. The phase diagram of a chiral magnet has some interesting features. For a particular region of the parameter space, the energy of single solitons is negative and a chiral soliton lattice (CSL), 
also called a spiral, is the ground state \cite{togawa2012chiral,KISHINE20151,PhysRevB.97.184303}, since one dimensional modulated states have a lower energy than skyrmions. There is another region where the energy of a single skyrmion is negative and the ground state becomes a lattice of skyrmions \cite{Lin:2014ada,Rossler:2006,Han:2010by,Ross:2020hsw} (See also Refs~\onlinecite{doi:10.1126/science.1166767,doi:10.1038/nature09124,doi:10.1038/nphys2045}). There is even a solvable point, where a critically coupled model exists that possess exact skyrmion solutions \cite{Barton-Singer:2018dlh}.  Finally, there are also ferromagnetic (FM) regions where the ground state is %ferromagnetic 
FM and skyrmions appear as positive energy solitons. Both isolated skyrmions \cite{doi:10.1126/science.1240573} and skyrmion tubes \cite{Wolf_2021} have been observed. 

Magnetic domain walls are another solitonic object that appears in chiral magnets with an easy-axis potential. They have also been a subject of particular study due to their application to magnetic memories \cite{doi:10.1126/science.1145799,KUMAR20221}.

Thus, topological aspects of (chiral) magnets have 
attracted much recent attention.
For instance, 
apart from domain walls and skyrmions, 
a lot of studies have been devoted to 
various topological objects 
such as 
monopoles \cite{tanigaki2015,fujishiro2019topological}, 
Hopfions \cite{Sutcliffe:2018vcb} 
and 
instantons \cite{Hongo:2019nfr}, 
see Ref.~\onlinecite{GOBEL20211} for a review.

Composite objects called domain-wall skyrmions\footnote{
The term 
``domain-wall skyrmions'' was first introduced in Ref.~\onlinecite{Eto:2005cc} 
in which Yang-Mills instantons 
in the bulk are 
3D skyrmions inside a domain wall. 
The terminology of this paper is different 
from Ref.~\onlinecite{Eto:2005cc}; what was studied there should be called 
domain-wall instantons 
in the current terminology.
}, formed by combining skyrmions with domain walls have been studied in 
quantum field theory \cite{Nitta:2012xq,Kobayashi:2013ju} (see also Refs.~\onlinecite{Jennings:2013aea,Bychkov:2016cwc})  
and more recently both theoretically \cite{PhysRevB.99.184412,PhysRevB.102.094402,KBRBSK,Ross:2022vsa,Amari:2023gqv} and experimentally \cite{Nagase:2020imn,li2021magnetic,Yang:2021} in chiral magnets  
(see also Refs.~\onlinecite{Kim:2017lsi,Lee:2022rxi}).
These are the two-dimensional counterparts of 
the three-dimensional domain-wall skyrmions in quantum field theory~\cite{Nitta:2012wi,Nitta:2012rq,
Gudnason:2014nba,Gudnason:2014hsa,Eto:2015uqa,Nitta:2022ahj}, with recent interests in application to QCD in a strong magnetic field 
\cite{Eto:2023lyo,Eto:2023wul} 
or rapid rotation~\cite{Eto:2023tuu}.
The trajectories of skyrmions under an applied current are bent in the bulk because of the skyrmion Hall effect, 
yielding difficulty when controlling the motion of skyrmions.
In contrast, skyrmions on a domain wall only move along the domain wall. Thus, domain-wall skyrmions 
are expected to be useful for constructing easily controllable magnetic memories. 
Thus far, domain-wall skyrmions have been studied in the %(anti-)ferromagnetic 
FM phase, but one important direction is to explore these objects 
in CSL phases, 
which is the main target of this paper.

In this paper, we use a mixture of analytic and numerical techniques to study chiral magnets with an out-of-plane easy-axis anisotropy term. We construct domain-wall skyrmion configurations in these systems, compare them to the earlier work on domain-wall skyrmions in chiral magnets, and study the stability of domain-wall skyrmions 
in the %(anti-)ferromagnetic 
FM phase and the CSL 
(spiral) phase.
It was previously found in  Refs.~\onlinecite{PhysRevB.99.184412,Amari:2023gqv} that the worldline of a domain-wall skyrmion is bent to form a cusp at the position of the skyrmion. 
Firstly, we describe such a cusp using both the analytic method of the moduli approximation,  
sometimes called the Manton approximation 
\cite{Manton:1981mp,Eto:2006pg,Eto:2006uw} 
(a double sine-Gordon equation in Ref.~\onlinecite{PhysRevB.99.184412})  
and a numerical simulation, 
and find a good match between them at least when the DM interaction is small.
We also show that 
the cusp grows toward the 
phase boundary with the CSL, 
and eventually diverges at 
the phase boundary, implying the instability of domain-wall skyrmions in the CSL.
Secondly, we numerically confirm that domain-wall skyrmions are in fact unstable in the CSL phase: a pair of a domain-wall skyrmion and an anti-domain wall without a skyrmion decays into a bimeron through a reconnection process. A bimeron in a CSL was previously studied in Ref.~\onlinecite{PhysRevB.106.224428} for the case of a Zeeman term coupling to a magnetic field and without easy-axis anisotropy, in which case a single soliton is cut into two pieces ending on merons. 
By contrast, in our case, a pair annihilation of the domain wall and anti-domain wall occurs locally, 
and they are connected by a U-shape at two positions.
Thirdly, we also present a construction of a chain of domain-wall skyrmions from an analytical approach making use of the double sine-Gordon equation and its solutions.\footnote{A three-dimensional version of a domain-wall skyrmion chain has been recently found in QCD \cite{Eto:2023wul}.
}
More precisely, we find that  when domain-wall skyrmions and domain-wall anti-skyrmions appear alternately in the CSL, 
the cusps of the (anti-)domain walls are in the same direction and the whole configuration is stable, 
but the (anti-)domain walls are bent logarithmically.

This paper is organized as follows. In Sec.~\ref{sec:model}, we review the model of a chiral magnet and revise some of its key features such as the domain wall and CSL. In Sec.~\ref{sec:single}, we present the effective theory for a domain-wall skyrmion, reviewing the details of Refs.~\onlinecite{PhysRevB.99.184412,Amari:2023gqv,Ross:2022vsa} and give a comparison between the theory and the numerics. Sec.~\ref{sec:instability}, studies a composite object consisting of a domain-wall skyrmion and a domain wall, and shows that such an object is unstable and decays into a bimeron. Then in Sec.~\ref{sec:chain}, we construct a chain of domain-wall skyrmions by considering a CSL with modulations on top of it. Finally in Sec.\ref{sec:discussion} we give a summary and discuss some open questions.

\section{Model and its ground states}
\label{sec:model}
\if0{
Contents:
\begin{itemize}
	\item Definition of the model  
	\item DW and CSL solution
	\item Properties of CSL (pitch, energy minimization condition)
\end{itemize}
\hrulefill
\newline
}\fi

In this section, we introduce the Hamiltonian relevant to the study of chiral magnets, and review the domain wall and CSL solutions supported by this Hamiltonian. We then briefly review the ground state of the model and some of the properties of the CSL that will be important in later sections.

Let $\n=(n_x,n_y,n_z)$ be a unit vector representing a magnetization vector. We consider a classical spin system on a square lattice defined by a Hamiltonian of the form
\begin{align}
 H=&\sum_{\langle i, j\rangle}\left[J \n\left(\vec{r}_{i}\right) \cdot \n\left(\vec{r}_{j}\right)+\D\left(\vec{r}_{i}, \vec{r}_{j}\right) \cdot\left\{\n\left(\vec{r}_{i}\right) \times \n\left(\vec{r}_{j}\right)\right\}\right] 
\notag\\
&-\frac{\mu^{2}}{2} \sum_{i} n_{z}^{2}\left(\vec{r}_{i}\right) \ ,
\end{align}
where the first term is the exchange interaction, the second is the DM interaction, and the third is an out-of-plane easy-axis anisotropy. We are here considering the exchange interaction to be %ferromagnetic
FM, i.e.,  $J<0$, and take a DM vector of the form 
\begin{equation}
\begin{aligned}
& \D\left(\vec{r}_{i}, \vec{r}_{i} \pm a \vec{e}_{x}\right) \equiv \pm \D_{x}=\mp \kappa(\cos \vartheta,-\sin \vartheta, 0) \ ,
\\
& 
\D\left(\vec{r}_{i}, \vec{r}_{i} \pm a \vec{e}_{y}\right) \equiv \pm \D_{y}=\mp \kappa(\sin \vartheta, \cos \vartheta, 0) \ ,
\end{aligned}
\end{equation}
where $a$ is the lattice constant, $\kappa$ and $\vartheta$ are constant.
In the continuum limit, this Hamiltonian can be written as
$    H=\int d^2x ~ {\cal H} + \const \ ,$
with
\begin{align}
\begin{aligned}
    {\cal H} =&
\frac{|J|}{2} \partial_{k} \n \cdot \partial_{k} \n
\\
&\quad
 +a^{-1} \D_{k} \cdot\left(\n \times \partial_{k} \n\right)
%+\frac{\kappa}{a}\left\{\cos\vartheta ~\n\cdot \nabla\times \n  -\sin\vartheta ~\n \cdot \left[\left(\vec{e}_z\times \nabla \right)\times \n \right] \right\}
+\frac{\mu^{2}}{2 a^{2}}\left(1-n_{z}^{2}\right) \ .
\end{aligned}
\label{Hdens_n}
\end{align}
The DM term can explicitly be written in the form
\begin{align}
    \begin{aligned}
        \D_{k} \cdot\left(\n \times \partial_{k} \n\right)=\kappa&
        \left\{\cos\vartheta ~\n\cdot \nabla\times \n 
        \right. \\
        &\left. \quad
        -\sin\vartheta ~\n \cdot \left[\left(\vec{e}_z\times \nabla \right)\times \n \right] \right\} \ .
    \end{aligned}
\end{align}
The angular parameter $\vartheta$ differentiates between different types of spin-orbit coupling in the underlying lattice spin system. The DM term with $\vartheta=0$ can arise from the Dresselhaus spin-orbit coupling, and that with $\vartheta=\pi/2$ corresponds to the Rashba spin-orbit coupling.
Hereafter, we use units where $|J|=a=1$ for simplicity.

In this paper, we focus on the case where domain walls are placed orthogonal to the $x$-axis. However, the generalization to the other cases is straightforward, thanks to the rotational symmetry of the Hamiltonian.
Under the spatial rotation
\begin{equation}
    \begin{pmatrix}
        x\\
        y
    \end{pmatrix}
    \to
    \begin{pmatrix}
        \tilde{x}\\
        \tilde{y}
    \end{pmatrix}
  =\begin{pmatrix}
      x\cos \gamma  +y\sin \gamma
      \\
      -x\sin \gamma  +y\cos \gamma 
  \end{pmatrix} \ ,
\end{equation}
all the terms in the Hamiltonian are invariant, except for the DM term. The DM vector transforms as 
\begin{align}
\begin{aligned}
     &\D_{x}\to\tilde{\D}_{x}=-\kappa(\cos \tilde{\vartheta},-\sin \tilde{\vartheta}, 0) ,
     \\
     &\D_y \to \tilde{\D}_{y}=-\kappa(\sin \tilde{\vartheta}, \cos \tilde{\vartheta}, 0)
\end{aligned}
\end{align}
with $\tilde{\vartheta}=\vartheta-\gamma$.
Therefore, when one wants to consider a domain wall normal to the $\tilde{x}$-axis, the results can be obtained by just replacing $\vartheta$ in the following formulae with $\tilde{\vartheta}$.

In order to describe domain wall and CSL solutions normal to the $x$-axis, we parametrize the magnetization vector as
\begin{equation}
  \n=(\cos \phi \sin f, \sin \phi \sin f, \cos f)  
    \label{n_parametrization}
\end{equation}
and employ an ansatz of the form
\begin{equation}
    f=f(x), \qquad \phi=\const \ .
      \label{ansatz_CSL}
\end{equation}
Substituting the ansatz \eqref{ansatz_CSL} into the Hamiltonian density \eqref{Hdens_n}, we obtain
\begin{equation}
    \calH=\frac{1}{2}\left(\partial_{x} f \right)^{2}+\kappa \sin (\vartheta+\phi) \partial_{x} f+\frac{\mu^{2}}{2} \sin^{2} f+\const \ ,
    \label{eq:H_chiralSG}
\end{equation}
which is the so-called chiral sine-Gordon model.
As its name implies, the Euler-Lagrange equation with respect to $f$ is given by the sine-Gordon equation
\begin{equation}
    \partial_{x}^{2} f=\frac{\mu^{2}}{2} \sin 2 f \ ,
\end{equation}
because the second term in the Hamiltonian \eqref{eq:H_chiralSG} is a total derivative term.
The single-domain wall solution is given by the sine-Gordon kink
\begin{equation}
    f=2 \arctan [\exp (\mu x+X)] \ ,
    \label{1-kink_profile}
\end{equation}
where $X$ is a moduli parameter relevant to the position of the domain wall. 
Note that this solution represents a $\pi$-domain wall, which means that $|f(+\infty)-f(-\infty)|=\pi$.
We call the solution with $\mu>0$ a domain wall and that with $\mu<0$ an anti-domain wall, when we distinguish them.
In addition to Eq.~\eqref{1-kink_profile}, the sine-Gordon equation possesses solutions describing a CSL. These are given in terms of the Jacobi amplitude function as
\begin{equation}
    f=\am\left(\frac{\mu}{\lambda} x+X, \lambda\right)+\frac{\pi}{2} \ ,
    \label{CSL_profile}
\end{equation}
where $\lambda \in(0,1]$ is the elliptic modulus. When $\lambda=1$, this reduces to the single kink solution \eqref{1-kink_profile}.
On the other hand, the Euler-Lagrange equation with respect to $\phi$ is just $\cos(\vartheta+\phi)=0$. The stable solution, which minimizes the Hamiltonian, is given by
\begin{equation}
    \phi=-\vartheta-\sign (\kappa \mu)  \frac{\pi}{2} +2l\pi \ ,
    \label{Eq:phi_field}
\end{equation}
with $l\in \integers$, because when $\mu$ is positive (negative), Eqs.~\eqref{1-kink_profile} and \eqref{CSL_profile} are monotonically increasing (decreasing) functions of $x$, i.e., $\sign(\mu) \pd_x f>0$.
Note that Eqs.~\eqref{1-kink_profile} and \eqref{CSL_profile} are solutions of this model for an arbitrary value of $\phi$, because $\phi$ does not appear in the equation of motion for $f$. However, it does contribute to the Hamiltonian through the total derivative term coming from the DM interaction. This means that $\phi$ is a quasi-moduli parameter. Note that when there is no DM term, $\phi$ is a true moduli parameter such as in the studies of domain-wall skyrmions in Refs.~\onlinecite{Nitta:2012xq,Kobayashi:2013ju}.

We now discuss some properties of the CSL solution. From the periodicity of the Jacobi amplitude function, one finds that the period of the CSL is given by
\begin{equation}
    L=\frac{4 \lambda}{|\mu|} \K(\lambda) \ ,
\end{equation}
where $\K(\lambda)$ is the elliptic integral of the first kind. The energy per unit length in the $y$-direction, $\int dx~ {\cal H}$, takes its minimum when $\lambda$ satisfies
\begin{equation}
   \frac{\E(\lambda)}{\lambda}=\left|\frac{\kappa}{\mu}\right| \frac{\pi}{2}  \ ,
   \label{condition_modulus}
\end{equation}
where $\E(\lambda)$ is the elliptic integral of the second kind.
The energy of the CSL becomes lower than that of a uniform vacuum state when $4\mu^2 < \kappa^2 \pi^2$.
In this parameter region, the ground state is a CSL phase. On the other hand, in the region $4\mu^2 > \kappa^2 \pi^2$, the ground state is a %ferromagnetic 
FM phase.

%%%%%%%%%%%%%%%%%%%%%%%%%%%%%%%%%%%%%%%%%%%%%%%%%%%%%%%%%%%%%
%\newpage
\section{Single Domain-wall skyrmion}
\label{sec:single}
\if0{
Contents:
\begin{itemize}
	\item Construction of the eff th. 
	\item Solution
	\item Comparision between eff th. and numerics
\end{itemize}
\hrulefill
\newline
}\fi

%%%%%%%%%%%%%%%%
\subsection{Effective theory approach}
\label{subsec:effective_theory}
This subsection is essentially a review of Ref.~\onlinecite{PhysRevB.99.184412}.
We study domain-wall skyrmions in the FM phase of the theory using the moduli  approximation 
(sometimes called the Manton approximation)
\cite{Manton:1981mp,Eto:2006pg,Eto:2006uw}. 
For this purpose, we consider the domain wall solution and promote the (quasi-)moduli parameters $\{X, \phi\}$ to fields depending on $y$, the coordinate along the domain wall, namely,
\begin{align}
& f=2 \arctan \left[\exp \left(\mu x+X\left(y\right)\right)\right] \ ,
\label{eq:approx_f}
\\
& \phi=\varphi\left(y\right)-\vartheta \ .
\label{eq:approx_phi}
\end{align}
Substituting these fields into the Hamiltonian density \eqref{Hdens_n}, we get an effective energy on the domain wall of the form
\begin{align}
{\cal E}_{\text {eff }}= & \int d x ~ \calH 
\notag \\
= & |\mu|^{-1}\left\{\left(\partial_{y} X \right)^{2}+\left(\partial_{y} \varphi\right)^{2} 
\right.
\notag\\
& \qquad~ +\kappa \pi\left( \mu \sin \varphi - \partial_{y} X \cos \varphi \right) \Big\}  + \const
\end{align}
where we have used $\arctan \xi+\arctan \xi^{-1}={\pi/2}$.
The Euler-Lagrange equations associated with this effective energy are given by\footnote{
In our previous work \cite{Ross:2022vsa}, we assumed 
$X=0$ for simplicity, 
but this is only justified when the domain wall tension is large. In such an approximation, 
we do not see the cusp structure discussed below.
}

\begin{align}
& \partial_{y}^{2} \varphi -
\Tkappa \left\{\mu \cos \varphi+\partial_{y} X \sin \varphi\right\}=0  \ ,
\label{eq:eom_phi}
\\
& \partial_{y}\left[\partial_{y} X-\Tkappa \cos \varphi\right]=0 \ ,
\label{eq:eom_X}
\end{align}
with $\Tkappa = \kappa\pi /2$.
It follows from Eq.~\eqref{eq:eom_X} that 
%\begin{align}
$
   \partial_{y} X=\Tkappa \cos \varphi+C, 
$
%\end{align}
for an arbitrary constant $C$.
Plugging it into the effective energy, one gets
\begin{align}
{\cal E}_{\text {eff }}
= & \frac{1}{|\mu|}\left[
\left(\partial_{y} \varphi\right)^{2}
+\left(\Tkappa\sin\varphi + \mu \right)^2 
+C^{2}\right] \ ,
\end{align}
where we omitted constant terms not depending on $C$.
Clearly, $C=0$ is energetically preferred. So, we let $C=0$ hereafter since we are interested in the lowest energy solitonic excitations on the domain wall. 
Then, the equations we want to solve read
\begin{align}
& \partial_{y}^{2} \varphi=\Tkappa\mu \cos \varphi+\frac{\Tkappa^{2}}{2} \sin (2 \varphi)  \ , 
\label{eq:DSG}\\
& \partial_{y} X=\Tkappa \cos \varphi \ ,
\label{eq:pdX}
\end{align}
where Eq.~\eqref{eq:DSG} is nothing but the double sine-Gordon equation. 
Solutions of the double sine-Gordon equation were studied 
for chiral magnets in the presence of both the 
easy-axis anisotropy and Zeeman magnetic field in Refs.~\onlinecite{PhysRevB.65.064433,Ross:2020orc}

%%%%%%%%%%%%%%%%%%%%%%%%%%%%%%%%%%%%%%%%%%%%%
\begin{figure*}[!ht]
    \centering
    \includegraphics[width=1.0\linewidth]{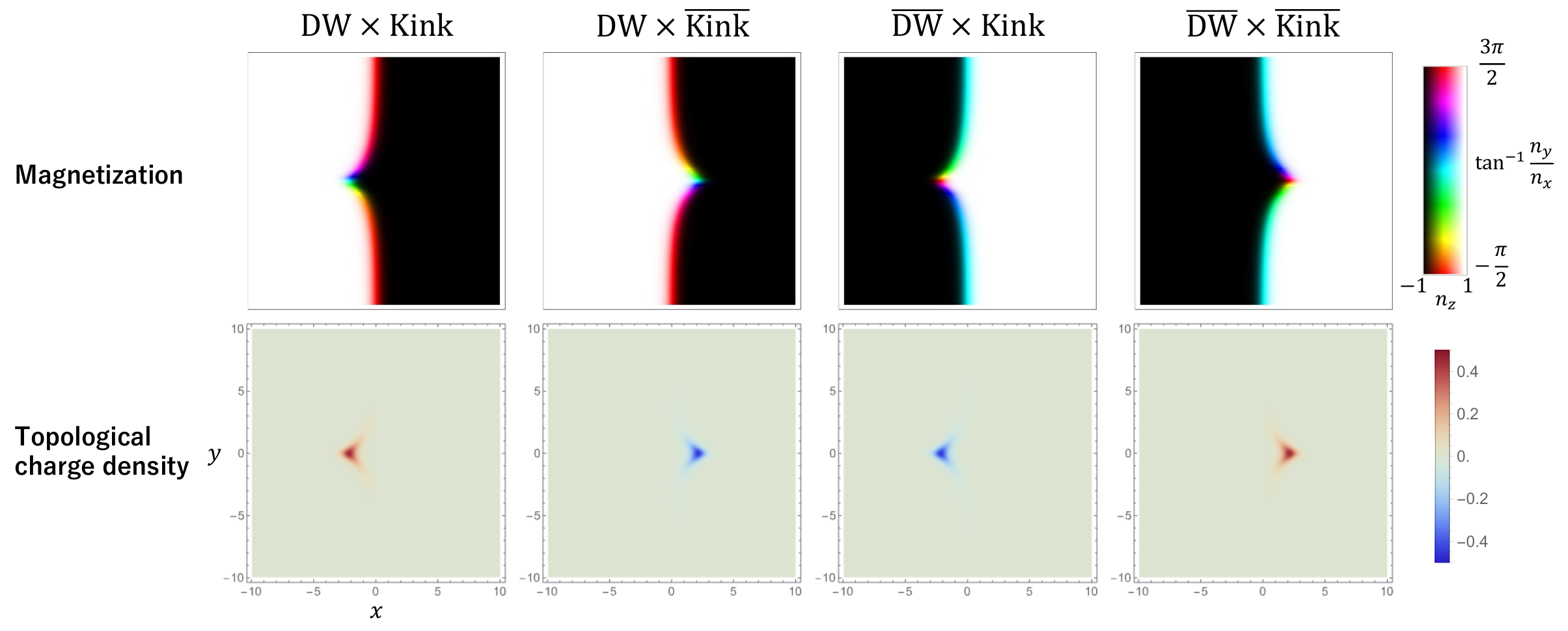}
    \caption{Single domain-wall skyrmions using the moduli approximation. The top panels represent the profile of the magnetization vector given with the solutions \eqref{sol_phi} and \eqref{sol_X}, and the bottom panels their topological charge density. For those figures, we used $\kappa=1.0, \mu^2=3$, and $\vartheta=\alpha=\beta=0$. The name of the configurations, e.g. DW$\times$Kink, specifies the sign of $\mu$ and the sign in Eq.~\eqref{sol_phi}, where $\overline{\text{DW}}$ and $\overline{\text{Kink}}$ denotes anti-DW and anti-kink, respectively.}
    \label{fig:DW-skyrmion_effth}
\end{figure*}
%%%%%%%%%%%%%%%%%%%%%%%%%%%%%%%%%%%%%%%%%%%%%

Since Eq.~\eqref{eq:DSG} does not contain $X$, we can solve the system of the equations by first finding a solution to Eq.~\eqref{eq:DSG} and then integrating Eq.~\eqref{eq:pdX} using this solution.
We consider a single kink solution of Eq.~\eqref{eq:DSG} which asymptotically decays into the lowest energy vacuum configuration on the domain wall \eqref{Eq:phi_field}.
To simplify the following calculation, we introduce
\begin{align}
    \psi = \varphi + \sign(\Tkappa\mu) \frac{\pi}{2} - 2 l \pi \ .
    \label{Eq:psi_def}
\end{align}
To take account of the asymptotic behavior of $\phi$, we impose boundary conditions for $\psi$ of the form
\begin{equation}
    \lim _{y \rightarrow - \infty} \partial_{y} \psi=0, 
    \quad\lim _{y \rightarrow - \infty} \cos \psi=1 \ .
    \label{eq:BC_phi}
\end{equation}
Eq.~\eqref{eq:DSG} with these boundary conditions implies the first-order differential equation
\begin{align}
 \partial_{y} \psi  
=\pm\sqrt{ 2|\Tkappa\mu| \left( 1- \cos \psi\right) -\Tkappa^2\left(1- \cos^{2} \psi  \right)} \ .
\end{align}
Note that the r.h.s is real because of the condition for the FM phase, $|\mu|>|\Tkappa|$.
It follows that 
\begin{widetext}
\begin{align}
 y&=\pm\int \frac{d \psi}{\sqrt{ 2|\Tkappa\mu| \left( 1- \cos \psi\right)-\Tkappa^2\left( 1 - \cos^{2} \psi  \right)}} 
 =\mp\frac{1}{\sqrt{|\Tkappa|(|\mu|-|\Tkappa|)}} \operatorname{arctanh} \sqrt{\frac{(|\mu|-|\Tkappa|)\cos^2\dfrac{\psi}{2}}{|\mu|-|\Tkappa|\cos^2\dfrac{\psi}{2} }} + \alpha
\end{align}
where $\alpha$ is a constant. Solving this equation for $\psi$ and using \eqref{Eq:psi_def}, we arrive at
\begin{align}
&\varphi=\pm 2 \arccos \left[
\frac{\sqrt{|\mu|}\sinh\left(\sqrt{|\Tkappa|(|\mu|-|\Tkappa|)}(y-\alpha) \right)}{\sqrt{|\mu|\cosh^2\left(\sqrt{|\Tkappa|(|\mu|-|\Tkappa|)}(y-\alpha) \right)-|\Tkappa|}} \right] -\sign(\Tkappa\mu) \frac{\pi}{2} + 2 l \pi \ .
    \label{eq:solution_phi}
\end{align}
The solution with the plus sign stands for an anti-kink, and the one with the minus sign gives a kink.
Substituting this solution into Eq.~\eqref{eq:pdX}, one gets
\begin{align}
\partial_{y} X =
\pm 2\sign(\mu)\frac{\sqrt{|\Tkappa\mu|}\sqrt{|\Tkappa|(|\mu| - |\Tkappa|)} \sinh \left(\sqrt{|\Tkappa|(|\mu|-|\Tkappa|)}(y-\alpha) \right)}{  |\mu|\cosh^2\left(\sqrt{|\Tkappa|(|\mu|-|\Tkappa|)}(y-\alpha) \right) - |\Tkappa|} \ .
\end{align}
Integrating the both sides directly, we obtain
\begin{equation}
    X=\mp 2
    \sign(\mu) \left( \operatorname{arctanh}\left[\sqrt{\frac{|\mu|}{|\Tkappa|}} \cosh \left(\sqrt{|\Tkappa|(|\mu|-|\Tkappa|)}(y-\alpha) \right)\right]
    +\frac{i\pi}{2}\right)
    +\beta  \ ,
    \label{eq:solution_X}
\end{equation}
\end{widetext}
where $\beta$ is a constant.

We have four types of domain-wall skyrmion corresponding to the combinations of (anti-)domain wall and (anti-)kink. In Fig.~\ref{fig:DW-skyrmion_effth}, we show the profile of the magnetization vector and the topological charge distribution for every type of domain-wall skyrmion, where the charge is defined as
\begin{align}
    Q=\frac{1}{4\pi}\int d^2x ~\n\cdot\left(\pd_x\n \times\pd_y\n \right) \ .
\end{align}
The topological charge of the solutions is either $\pm 1$. 
Note that as discussed in Ref.~\onlinecite{Ross:2022vsa}, the four solutions are all degenerate and stable, in contrast to isolated skyrmion and anti-skyrmion configurations above a %ferromagnetic 
FM background where only either skyrmions or anti-skyrmions can stably exist depending on the magnetic material. 
%\txtr{(CR: I feel like this sentence needs reworded but not sure how best to do this. The key point is that all the types of domain wall (anti-)skyrmions are stable while isolated skyrmions and anti-skyrmions are not both stable, right?; YA: Yes, it is.  )}
As can be seen in Fig.~\ref{fig:DW-skyrmion_effth}, the domain wall bends in a dogleg shape and the topological charge localizes near the turning point. Therefore, the smooth cusp of the domain wall can be viewed as the domain-wall skyrmion. 

\begin{figure}[!t]
    \centering
    \includegraphics[width=1.0\linewidth]{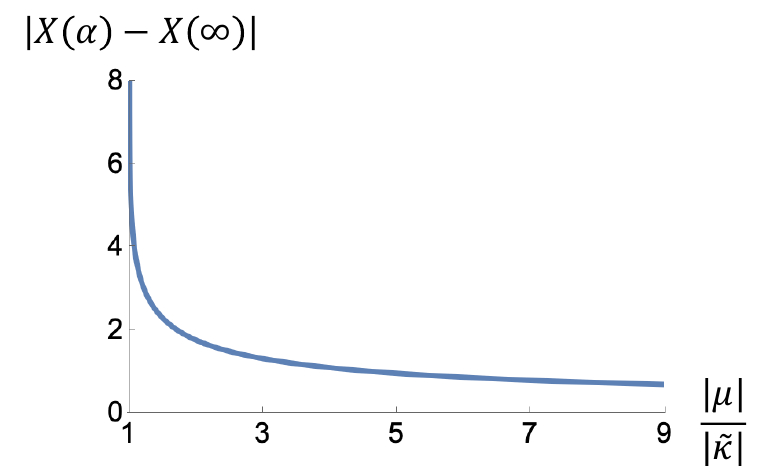}
    \caption{The domain-wall skyrmion position measured from the domain wall itself as a function of $|\mu|/|\Tkappa|$}
    \label{fig:X0}
\end{figure}

The bending of the domain wall is because of the non-triviality of $X$. The solution for $X$ also has inversion symmetry about $y=\alpha$. Moreover, the quantity $|X(y)-X(\pm\infty)|$ takes its maximal value at $y=\alpha$. 
One can identify $|X(\alpha)-X(\pm\infty)|$ as the position of the domain-wall skyrmion measured relative to the domain wall itself, and it has the following limits
\begin{align}
\begin{split}
   & \lim_{\frac{|\mu|}{|\Tkappa|}\to 1} |X(\alpha)-X(\pm\infty)|\to \infty \ ,
    \\
   & \lim_{\frac{|\mu|}{|\Tkappa|}\to \infty} |X(\alpha)-X(\pm\infty)|\to 0 \ .
\end{split}
\end{align}
This indicates that at the transition point between the FM and CSL phase, i.e., $|\mu|=|\Tkappa|$, the cusp stretches infinitely far away from the domain wall itself.
We plot $|X(\alpha)-X(\pm\infty)|$ as a function of $|\mu|/|\Tkappa|$ in Fig. \ref{fig:X0}.

%%%%%%%%%%%%%%%%%%%%%%%%%%%%%%%%%%%%%%%%%%%%%%%%%%%%%%
%%%%%%%%%%%%%%%%%%%%%%%%%%%%%%%%%%%%%%%%%%%%%%%%%%%%%%

%%%%%%%%%%%%%%%%
\subsection{Numerical solutions}
\if0{
%%%%%%%%%%% 
\begin{figure}[!t]
\centering
     \includegraphics[width=1\columnwidth]{figures/DW-skyrmion_BC.pdf} 
    \caption{Boundary condition used in the numerical simulation for the domain-wall skyrmions with the parameter $\kappa > 0, \vartheta=0 $.}
    \label{fig:DW-skyrmion_BC}
\end{figure}
%%%%%%%%%%%%%%
}\fi
%%%%%%%%%%%%%%%%%%%%%%%%%%%%%%%%%%%%%%%%%%%%%%%%%%%%%%
\begin{figure*}[!tp]
   \centering
    \includegraphics[width=0.8\linewidth]{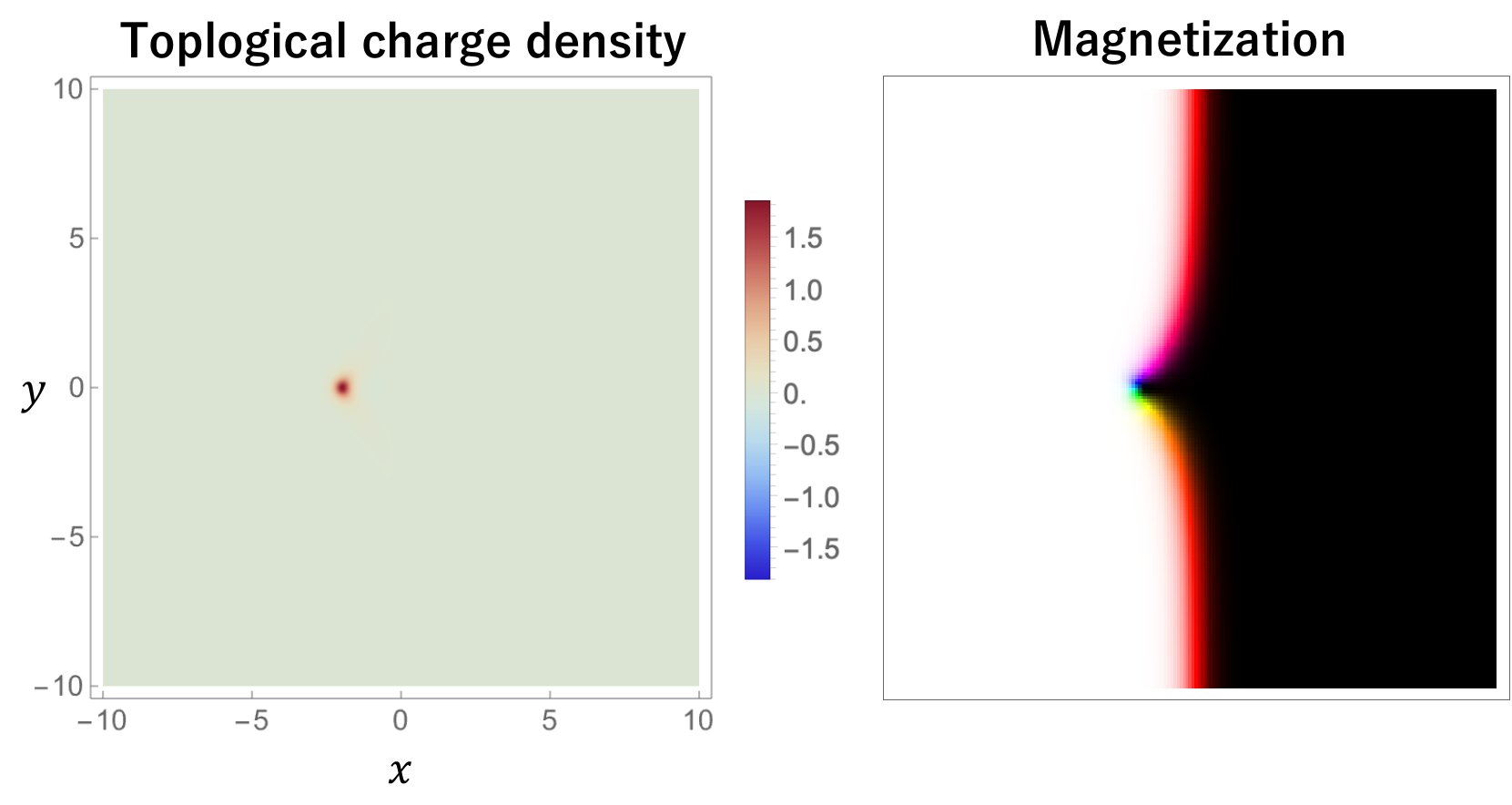}
    \caption{Numerical solution of a domain-wall skyrmion, $Q=1$ skyrmion trapped in a domain wall, corresponding to the left-most configuration, i.e., DW $\times$ Kink in Fig.~\ref{fig:DW-skyrmion_effth}. The left figure shows the topological charge density and the right represents the magnetization profile. We used for the simulation $\kappa=1.0, \mu=+\sqrt{3}$ and $\vartheta=0$.
    %\txtr{$\kappa$ for this simulation may be too large for the moduli approximation to work not only qualitatively but also quantitatively.}
    }
    \label{fig:DW-skyrmion}
\end{figure*}
%%%%%%%%%%%%%%%%%%%%%%%%%%%%%%%%%%%%%%%%%%%%%%%%%%%%%%

%%%%%%%%%%%%%%%%%%%%%%%%%%%%%%%%%%%%%%%%%%%%%%%%%%%%%%
\begin{figure*}[!htp]
   \centering
    \includegraphics[width=0.8\linewidth]{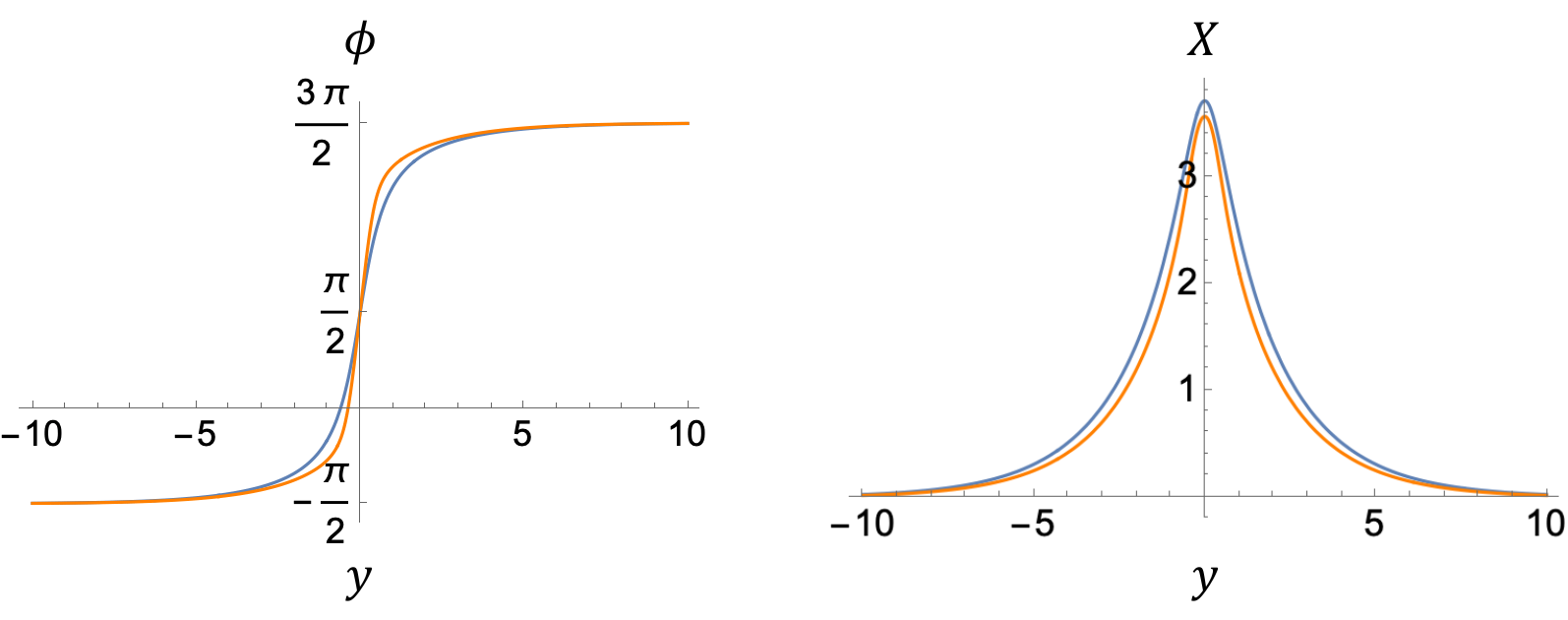}
    \caption{Comparision of the numerical solution of the system \eqref{Hdens_n} (Orange) and analytic one in the effective theory (Blue)  for $\phi$ and $X$.  The parameters used are $\kappa=1.0, \mu=+\sqrt{3}$. }
    \label{fig:DW-skyrmion_profile}
\end{figure*}
%%%%%%%%%%%%%%%%%%%%%%%%%%%%%%%%%%%%%%%%%%%%%%%%%%%%%%

%[Direction]
We numerically construct single domain-wall skyrmions in the model described by the Hamiltonian \eqref{Hdens_n} and compare them with the analytic solutions obtained in the previous subsection.
As we shall see, the numerical and analytic solutions in the moduli approximation exhibit good qualitative agreement.

%[Numerical setup]
The Euler-Lagrange equation associated with the Hamiltonian \eqref{Hdens_n} is given by
\begin{align}
        \partial_b^2 n_a 
    &+2\kappa\left[\cos\vartheta ~\varepsilon_{abc}\partial_b n_c
    +\sin\vartheta (\partial_a n_z - \delta_{az}\partial_bn_b) \right]
    \notag\\
    &+\mu^2\delta_{az}n_z + \Lambda n_a = 0
        \label{eq:eom_skyrmion}
\end{align}
where $\Lambda$ is a Lagrange multiplier.
We solve the equations of motion \eqref{eq:eom_skyrmion}
using a nonlinear conjugate gradient method with a finite difference approximation of fourth order, where we run our simulation on a grid with $401\times 401$ lattice points and lattice spacing $\Delta=0.1$. 
The initial input we used is given by the single domain wall solution for the easy-axis case in Eq.~\eqref{1-kink_profile} with the moduli parameter $X=0$ and 
\begin{equation}
    \phi = 4\arctan e^{c y} - \vartheta -\sign(\kappa\mu) \frac{\pi}{2}
\end{equation}
with a real parameter $c$. 
 We impose the Dirichlet boundary conditions relevant to the domain-wall skyrmions: we assign either $\n=(0,0,1)$ or $\n=(0,0,-1)$ to the boundaries in the $x$-direction, respectively, as they are compatible with the initial input; for the boundaries in the $y$-direction, the boundary value is fixed by the lowest energy domain wall solution given by Eqs.~\eqref{1-kink_profile} and \eqref{Eq:phi_field}. 

%[Results \& comparison]
The numerical solution is shown in Fig.~\ref{fig:DW-skyrmion}. 
One can observe that the numerical solution and the analytic solution in the effective theory are at least qualitatively in good agreement. 
Note that when $|\mu/\kappa|$ is smaller, the moduli approximation is quantitatively more accurate.
We compare the analytic solutions for $\{\phi, X\}$ and corresponding numerical data in Fig.~\ref{fig:DW-skyrmion_profile}. In order to extract data for $X$, we employ spline interpolation on the field $\mathbf{n}$. We then identify $X/\mu$ with $x$ where $n_z = 0$ is satisfied. The phase $\phi$ in Fig.~\ref{fig:DW-skyrmion_profile} is given by $\arctan(n_y/n_x)$ on the line $n_z = 0$.
We observe good agreement for both $\phi$ and $X$ between the numerical and analytic solutions.

%%%%%%%%%%%%%%%%%%%%%%%%%%%%%%%%%%%%%%%%%%%%%%%%%%%%%%%%%%%%%
%\newpage
\section{Instability of domain-wall skyrmions in the chiral soliton lattice phase, 
and domain-wall merons
}
\label{sec:instability}

%%%%%%%%%%%%%%%%%%%%%%%%%%%%%%%%%%%%%%%%%%%%%%%%%%%%%%
\begin{figure*}[!htp]
   \centering
    \includegraphics[width=1.0\linewidth]{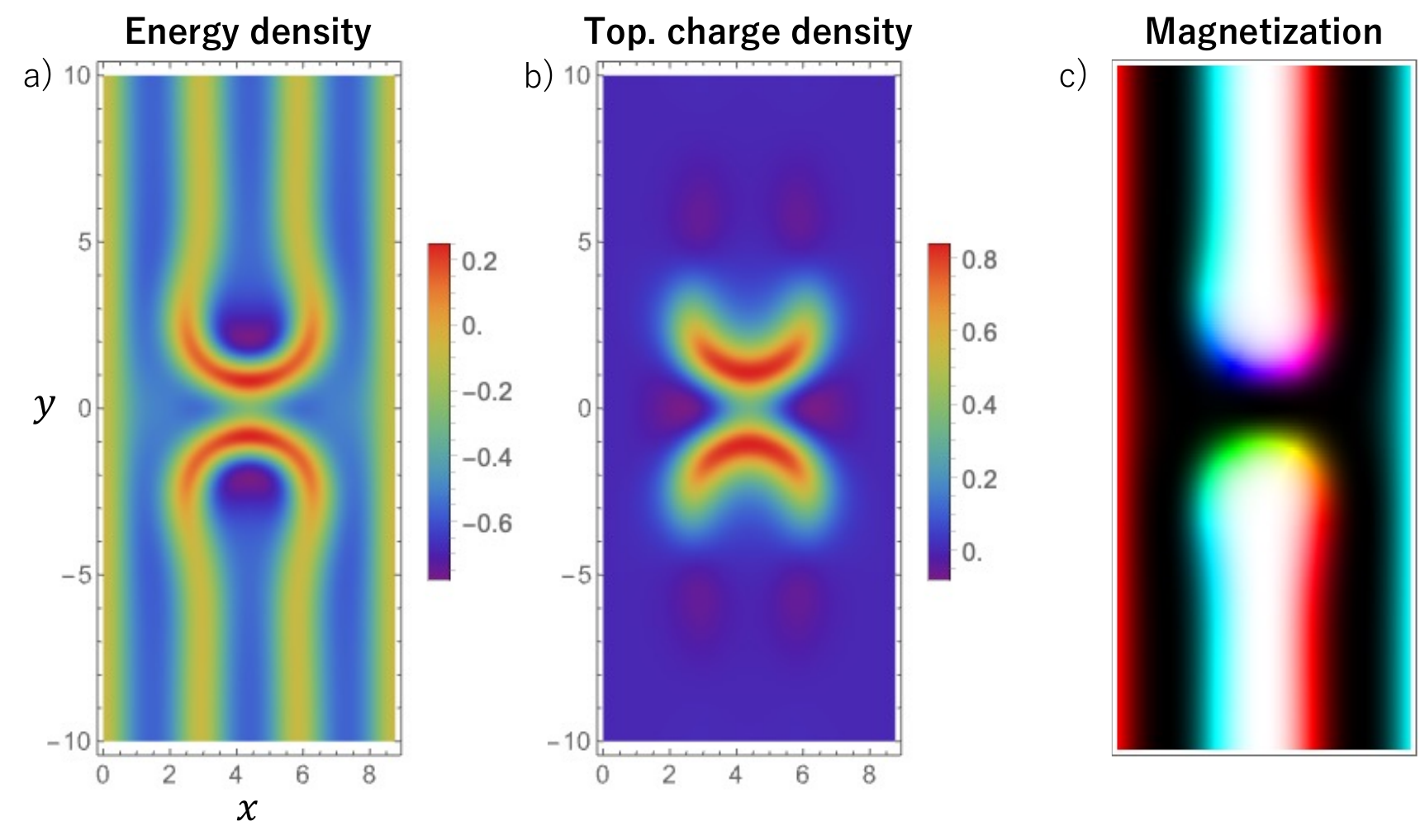}
    \caption{A bimeron in the CSL phase. We used the parameters $\kappa=1.1, \mu = 1.0$, and $\vartheta=0.0$ for the numerical simulation. Panel a) shows the energy density, b) the topological charge density, and c) the magnetization vector with the same color code as Fig.~\ref{fig:DW-skyrmion_effth}. We can recognize two lumps with a symmetric form in panel b). Since the total topological charge is unity, each lump should possess the charge of 1/2. }
    \label{fig:bimeron}
\end{figure*}
%%%%%%%%%%%%%%%%%%%%%%%%%%%%%%%%%%%%%%%%%%%%%%%%%%%%%%

%[Direction] 
So far, we have studied single domain-wall skyrmions in the FM phase. As discussed in Sec.~\ref{subsec:effective_theory}, the cusp of the domain wall corresponding to a domain-wall skyrmion infinitely extends from the domain wall itself at the FM-CSL phase boundary. 
Then, a natural question arises: can domain-wall skyrmions exist in the CSL phase, where domain walls are periodically arrayed and a cusp cannot extend to infinity?
In this section, we address the (in)stability of single domain-wall skyrmions in the CSL phase.
For simplicity, we restrict ourselves to the case with $\mu>0$.

%[Numerical setup]
To examine the (in)stability, we numerically solve the equation of motion \eqref{eq:eom_skyrmion} with an initial configuration describing a single domain-wall skyrmion in the CSL phase.
We prepare such configuration as follows.
For the function $f$, we use the CSL solution \eqref{CSL_profile} with $X=0$. 
The domain of the variable $x$ used for the simulations ranges from $0$ to $3L/2$. It follows that the domain walls are placed at $x=0$ and $x=L$, while anti-domain walls are located at $x=L/2$ and $x=3L/2$.
As for the phase $\phi$, we assign the anti-kink solution given in Eq.~\eqref{sol_phi} to the domain $x=[L/4,3L/4]$, and for the other domain, we employ the vacuum value \eqref{Eq:phi_field}.
This configuration represents a domain-wall skyrmion, $Q=1$ skyrmion trapped in an anti-domain wall, in the CSL.
The boundary condition we impose for $x=0$ and $x=3L/2$ are respectively $\n(0,y)=(0,1,0)$ and $\n(3L/2,y)=(0,-1,0)$.
We also consider a finite domain for $y$ and fix the field value at the boundary as the stable CSL solution given by Eqs.~\eqref{CSL_profile} and \eqref{Eq:phi_field}.

%[Results]
We apply a nonlinear conjugate gradient method with finite difference approximation for solving the equations of motion \eqref{eq:eom_skyrmion}.
We discretize the system into $400$ lattice sites along the $y$-axis with the lattice spacing $\Delta y=0.1$ and 150 sites along the $x$-axis.
In Fig.~\ref{fig:bimeron}, we show the resulting relaxed state. Note that we only plot the central 200 sites in the computational domain along the y-axis because the configuration in the excluded region is just a CSL. In this sense, the domain is large enough, and a boundary effect should be almost free for the configuration in Fig.~\ref{fig:bimeron}. 
One observes from the figures that the domain wall trapping a skyrmion and its neighboring domain wall reconnect and form a bimeron.
This result indicates that single domain-wall skyrmions in the CSL are unstable.

It is worth noting that 
 a bimeron in a CSL was previously studied in Ref.~\onlinecite{PhysRevB.106.224428} in the case of a Zeeman term coupling to a magnetic field and without easy-axis anisotropy. In that case, a CSL consists of sine-Gordon solitons rather than (anti-)domain walls, and a single soliton is cut into two pieces ending on merons. While the identifications of (anti-)domain walls or solitons are different between the case considered here and that considered in Ref.~\onlinecite{PhysRevB.106.224428}, the spin textures are topologically identical.

%%%%%%%%%%%%%%%%%%%%%%%%%%%%%%%%%%%%%%%%%%%%%%%%%%%%%%%%%%%%%
%\newpage
\section{Domain-wall skyrmion chain}
\label{sec:chain}

%%%%%%%%%%%%%%%%%%%%%%%%%%%%%%%%%%%%%%%%%%%%%%%%%%%%%%%%%%%%%
\begin{figure*}
    \centering
    \includegraphics[width=\textwidth]{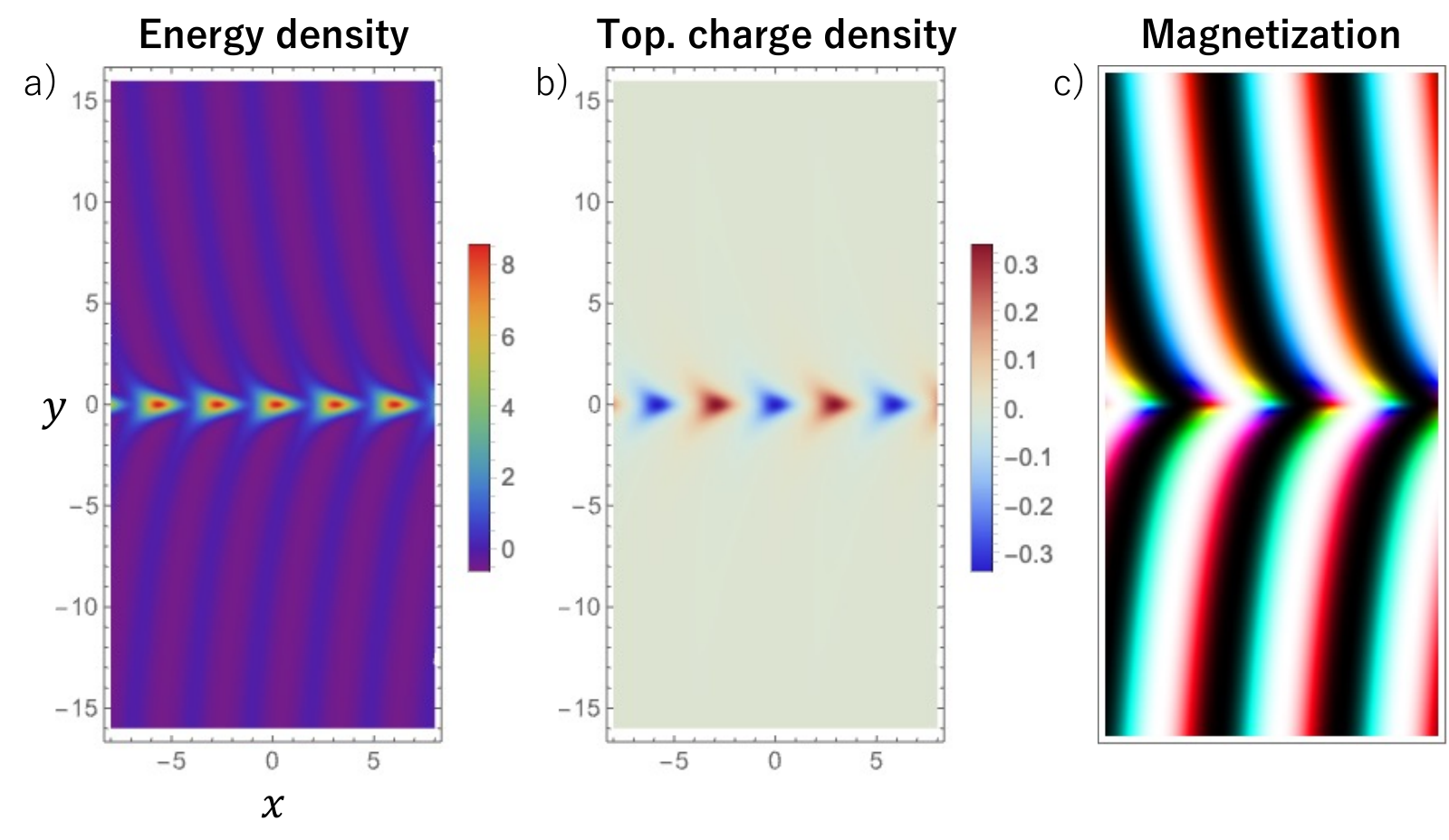}
    \caption{Quantities of the domain-wall skyrmion lattice solution with $\vartheta=0, \mu=1.0, \text{and } \kappa=1.1 $. Panel a) shows the energy density, b) topological charge density, and c) the profile of the magnetization vector with the same color code as Fig. \ref{fig:DW-skyrmion_effth}.We chose the plus sign in Eq.~\eqref{sol_phi} and \eqref{sol_X}, and let $\alpha=\beta=0$. }
    \label{fig:DWSkX_moduli}
\end{figure*}
%%%%%%%%%%%%%%%%%%%%%%%%%%%%%%%%%%%%%%%%%%%%%%%%%%%%%%%%%%%%%

In the previous section, we have shown that single domain-wall skyrmions are unstable in the CSL phase.
However, a configuration where domain-wall skyrmions are alternately arrayed with anti-domain-wall skyrmions, like DW $\times$ Kink and $\overline{\text{DW}} \times $Kink, can be meta-stable because the cusps may not crash into a neighboring wall. We call such a configuration a domain-wall skyrmion chain. Domain-wall skyrmion chains can be constructed using the moduli approximation, as we show in this section. 

The procedure to construct domain-wall skyrmion chains is parallel to that used for the single domain-wall skyrmion given in Sec. \ref{subsec:effective_theory}. The main difference is to utilize the function
\begin{align}
& f=\operatorname{am}\left(\frac{\mu}{\lambda} x+X(y), \lambda\right)+\frac{\pi}{2} \ , 
\end{align}
which represents a CSL with modulations, instead of Eq.~\eqref{eq:approx_f}. Note that for the phase $\phi$, we here use Eq.~\eqref{eq:approx_phi}, too.
The effective energy of the CSL per unit length can be defined as
\begin{align}
{\cal E}_\text{eff}  &=L^{-1} \int_{0}^{L} dx ~ {\cal H}
\notag \\
&=\frac{2}{L|\mu|}  \left[
\lambda^{-1}\left\{\E(\lambda)-\left(1-\lambda^{2}\right) \K(\lambda)\right\}\left(\partial_{y} \varphi\right)^{2} 
\right.
\notag \\
&\quad\qquad\qquad
+\lambda \E(\lambda)\left(\partial_y X - \frac{\Tkappa}{\E(\lambda)}\cos\varphi \right)^2
\notag\\
& \quad\qquad\qquad 
+2\Tkappa\mu\sin\varphi-|\Tkappa\mu|\cos^2\varphi{\Big]}
+ \const
\end{align}
where we used
\begin{equation}
    \am(4 \K(\lambda)+X, \lambda)=\am(X, \lambda) + 2 \pi \ .
\end{equation}
The equations that the lowest energy excitation should satisfy are given by
\begin{align}
&\partial_y^2\varphi =\frac{\lambda}{2}\frac{2\Tkappa\mu\cos\varphi + |\Tkappa\mu|\sin(2\varphi)}{\E(\lambda)-(1-\lambda^2)\K(\lambda)} \ ,
\label{eq:eom_phi_chain}
\\
& \partial_yX  =\frac{\Tkappa}{\E(\lambda)}\cos\varphi \ ,
\label{eq:eom_X_chain}
\end{align}
where Eq.~\eqref{eq:eom_phi_chain} can be obtained from the Euler-Lagrange equation with respect to $\varphi$ after using Eq.~\eqref{eq:eom_X_chain}.

Since Eq.~\eqref{eq:eom_phi_chain} does not include $X$, we first derive its solution and then solve Eq.~\eqref{eq:eom_X_chain}, similar to the approach in Sec.~\ref{subsec:effective_theory}.
We introduce $\psi$ through Eq.~\eqref{Eq:psi_def} and then the equation for $\varphi$ can be cast into the form
\begin{equation}
    \partial_{y}^{2} \psi=A^{2}\left(\sin \psi-\frac{1}{2} \sin (2 \psi)\right)
    \label{DSGeq}
\end{equation}
with
\begin{equation}
    A^{2}=\frac{\lambda|\Tkappa \mu| }{\E(\lambda)-\left(1-\lambda^{2}\right) \K(\lambda)}>0 \ .
\end{equation}
Note that the double sine-Gordon equation of this type \eqref{DSGeq} corresponds to the boundary of phase I and II in Ref.~\onlinecite{condat1983double}.
It follows from Eq.~\eqref{DSGeq} that 
\begin{align}
\frac{1}{2}\left(\partial_{y} \psi\right)^{2}=-A^{2}\left(\cos \psi+\frac{1}{2} \sin ^{2} \psi\right)+\const
\label{conservation_mechanical-energy}
\end{align}
%\txtr{(CR: At the moment \eqref{conservation_mechanical-energy} does not differentiate to give \eqref{DSGeq}, they differ by a sign in the $\sin(2\psi)$ term. I need to check (36) again as from comparing with Ref.~\onlinecite{condat1983double} they have a different term in the potential.)(YA: Thank you very much for pointing out my mistake. You are right. The sign in the $\sin(2\psi)$ term should be minus, and I corrected it.)(CR: that alll seems correct to me now, thanks for fixing that) }
Imposing the boundary condition Eq.~\eqref{eq:BC_phi},
we can write Eq.~\eqref{conservation_mechanical-energy} as
\begin{align}
\left(\partial_{y} \psi\right)^{2} 
 =A^{2}\left(1-\cos \psi\right)^2 \ .
\end{align}
Therefore, one obtains
\begin{align}
  y &= \pm \frac{1}{|A|}\int \frac{d \psi }{1-\cos \psi}
=\mp \frac{1}{|A|}\cot\frac{\psi}{2}+\alpha \ ,
\end{align}
where $\alpha$ is a moduli parameter.
Solving it inversely, we find 
\begin{equation}
    \varphi=\pm 2 \arccot\left[|A|(y - \alpha)\right] -\sign(\kappa\mu)\frac{\pi}{2} +2l\pi \ .
    \label{sol_phi}
\end{equation} 
Moreover, substituting the solution \eqref{sol_phi} into Eq.~\eqref{eq:eom_X_chain}, we get
\begin{equation}
    \partial_{y} X=\pm \frac{\mu}{\lambda} \frac{2|A|(y - \alpha)}{1+A^2(y - \alpha)^{2}}
\end{equation}
By integrating the both sides directly, we obtain
\begin{equation}
    X=\pm \frac{\mu}{\lambda} \frac{1}{|A|} \log \left[1+A^2(y - \alpha)^{2}\right]+\beta 
    \label{sol_X}
\end{equation}
where $\beta$ is a constant. 
%\txtr{(CR: I think that both (46) and \eqref{sol_X} should have a $1/\pi$ included if we are using \eqref{condition_modulus} to eliminate $E(\lambda)$. Does that make sense?)(YA: Yes, I used \eqref{condition_modulus}, but I think \eqref{sol_X} is correct because the coefficient in \eqref{eq:eom_X_chain} is not $\kappa/\E$ but $\Tkappa/\E=\sign(\kappa\mu)\lambda/\mu)$ )  (CR: You are right thanks for checking this.)}

Figure~\ref{fig:DWSkX_moduli} illustrates the domain-wall skyrmion chain that is anti-kinks on each domain wall in a CSL. As one easily sees from Eq.~\eqref{sol_X}, $X$ logarithmically diverges for large $|y|$. Therefore, it is difficult to obtain such a configuration by numerically solving the equations of motion \eqref{eq:eom_skyrmion} and to confirm their stability. However, we believe that domain-wall skyrmion chains exist as meta-stable solutions in the system \eqref{Hdens_n} and are well described by our solution, at least from the qualitative point of view.

%%%%%%%%%%%%%%%%%%%%%%%%%%%%%%%%%%%%%%%%%%%%%%%%%%%%%%%%%%%%%
%\newpage
\section{Conclusion and discussion}
\label{sec:discussion}

In this paper, we have explored the stability of 
domain-wall skyrmions 
in the %(anti-)ferromagnetic 
FM phase and the CSL 
(spiral) phase of chiral magnets with 
an out-of-plane easy-axis potential 
and without a Zeeman term.
We have studied the shape of the cusp 
of a domain-wall skyrmion
 in the %(anti-)ferromagnetic 
 FM phase 
 by using both an analytic method 
(in this case the double sine-Gordon equation) 
and numerical simulations 
based on the relaxation method. There is a good agreement between the two approaches as shown in Fig.~\ref{fig:DW-skyrmion_profile}.
Next, we have shown that 
the cusp grows as we approach the 
boundary with the CSL phase, 
and eventually diverges at 
the phase boundary 
as shown in Fig.~\ref{fig:X0}, implying the instability of domain-wall skyrmions in the CSL phase.
We have confirmed this instability by numerically showing that a pair of a domain-wall skyrmion and an anti-domain wall without a skyrmion decays into a bimeron through a reconnection process and are connected through the U-shapes 
between the two regions as in Fig.~\ref{fig:bimeron}.
Finally, we found that if domain-wall skyrmions and domain-wall anti-skyrmions appear alternately in the CSL phase,  
the configuration is stable because the directions of the cusps are the same, 
as shown in Fig.~\ref{fig:DWSkX_moduli}.
The worldline of each (anti-)domain wall is logarithmically bent 
away from the $y$ position of skyrmions as in Eq.~(\ref{sol_X}).

Before closing this paper, here we address some discussions.
In this paper, we have explored fundamental tools to study domain-wall skyrmions, 
and established the validity of analytic methods such as the moduli approximation. 
The next step is to apply these methods to construct more realistic magnetic memories. This could be done by carrying out simulations of race track memory for a system of domain-wall skyrmions.

The logarithmic bending of the shape of (anti-)domain walls in the skyrmion chain
could be avoided if we put the opposite 
structure in a different position 
of the $y$ coordinate.
In this case, there is a pair of 
a kink and anti-kink on each (anti-)domain wall, 
and so the topological lump charge is zero on each domain wall. 
Thus, the worldline of each soliton is asymptotically flat in the $y$ direction.\footnote{
This situation is similar to the so-called D-brane solitons in the
O(3) model with easy-plane anisotropy in three dimensional space 
(without the DM interaction)
\cite{Gauntlett:2000de,Isozumi:2004vg,Eto:2008mf}. 
In three dimensions, skyrmions are lines (strings) and 
a domain wall has a two-dimensional surface. 
A D-brane soliton consists of skyrmion lines that end on the domain wall from its both sides. 
If the numbers of skyrmions ending on the wall from one side and those ending on the other side are imbalanced, 
the domain wall is bent logarithmically, and 
if balanced, it is asymptotically flat. 
Although the dimensionality differs by one from that considered here, they may be related by 
a dimensional reduction.
}

It is known in field theory that a periodic array of a skyrmion with a twisted boundary condition reduces to a meron pair and 
eventually becomes a domain wall in the small periodicity limit 
\cite{Eto:2004rz,Eto:2006mz,Eto:2006pg}. 
In such a case, 
each meron is in fact a (anti-)domain wall 
where the $U(1)$ phase is twisted half along its worldline. 
This meron is topologically the same as the meron with the U-shape found in this paper.

Finally, in our previous paper \cite{Amari:2023gqv}, 
we gave a string theory realization of chiral magnets, 
where magnetic domain walls, skyrmions, and domain-wall skyrmions are represented by D-branes in string theory. 
All the configurations discussed in this paper, 
such as bimerons and domain-wall skyrmion chains 
can be realized by D-brane configurations. 
Investigating impacts of such configurations on string theory remains a future problem.

%%%%%%%%%%%%%%%%%%%%%%%%%%%%%%%%%%%%%%%%%%%%%%%%%%%%%%%%%%%%%
%\newpage

\begin{acknowledgments}

This work is supported in part by 
 JSPS KAKENHI [Grants No. JP23KJ1881 (YA) and No. JP22H01221 (MN)], the WPI program ``Sustainability with Knotted Chiral Meta Matter (SKCM$^2$)'' at Hiroshima University.
 The numerical computations in this paper were run on the ``GOVORUN" cluster supported by the LIT, JINR.
 
\end{acknowledgments}

%%%%%%%%%%%%%%%%%%%%%%%%%%%%%%%%%%%%%%%%%%%%%%%%%%%%%%%%%%%%%
\if0{
\newpage
\begin{appendix}
\section{Kink solution in the double sine-Gordon equation}
 \label{sec:DSG}
\setcounter{equation}{0}
\renewcommand{\theequation}{A.\arabic{equation}}

\end{appendix}
}\fi

%\bibliographystyleSM{apsrev4-1}
%\bibliographySM{refs}
%\bibliographystyle{apsrev4-1}
%\bibliography{refs}


\end{document}